\documentclass{elsart}
\usepackage{graphicx}
\usepackage{amsmath}
\usepackage{amssymb}
\newcommand{\GeVc}    {\mbox{$ {\mathrm{GeV}}/c                            $}}
\newcommand{\GeVcc}    {\mbox{$ {\mathrm{GeV}}/c^2                           $}}
\newcommand{\MeVc}    {\mbox{$ {\mathrm{MeV}}/c                            $}}

\newcommand{\mydeg}   {\mbox{$ ^\circ                                      $}}

\newcommand{\cher}{\u{C}erenkov }
\newcommand{\cherr}{\u{C}erenkov}

\newcommand{\micron}{\mbox{$\mu {\rm m}$}}
\def\NIMA#1#2#3{{\rm Nucl.~Instr.~and~Meth.} {\bf{A#1}} (#2) #3}

\def\PLB{{\em Phys. Lett.}  B}

\def\PLB#1#2#3{{\rm Phys.~Lett.} {\bf{B#1}} (#2) #3}
\begin{document}
\begin{frontmatter}
\title{The AMS-01 Aerogel Threshold \cher counter.}

\author[isn]{D. Barancourt},
\author[lapp,lip]{F. Barao},
\author[isn]{G. Barbier},
\author[lip]{G. Barreira},
\author[isn]{M. Bu\'enerd},
\author[fir]{G. Castellini},
\author[mos]{E. Choumilov},
\author[lapp]{J. Favier},
\author[lapp]{N. Fouque},
\author[tai]{A. Gougas},
\author[lapp]{V. Hermel},
\author[lapp]{R. Kossakowski},
\author[isn]{G. Laborie},
\author[bol]{G. Laurenti},
\author[tai]{S.-C. Lee},
\author[isn]{F. Mayet\thanksref{corr}},
\author[isn]{B. Meillon},
\author[tai]{Y.-T. Oyang},
\author[mos]{V. Plyaskin},
\author[mos]{V. Pojidaev},
\author[isn]{C. Rossin},
\author[isn]{D. Santos},
\author[isn]{F. Vezzu},
\author[lapp]{J.P. Vialle}
\thanks[corr]{corresponding author : Frederic.Mayet@isn.in2p3.fr (phone: +33 4-76-28-40-21, fax: +33 4-76-28-40-04)}
\address[isn]{Institut des Sciences Nucl\'eaires, 
 CNRS/IN2P3 and Universit\'e Joseph Fourier, 
 53, avenue des Martyrs, 38026 Grenoble cedex, France}
\address[lapp]{Laboratoire d'Annecy-le-Vieux de Physique des Particules (IN2P3-CNRS)   
                                F-74941 Annecy-le-Vieux cedex, France}
 
\address[lip]{Laborat\'orio de Instrumenta\c{c}\~{a}o e F\'{\i}sica Experimental de Part\'{\i}culas, 
          Avenida Elias Garcia, 14-1., 1000 Lisboa, Portugal}
\address[fir]{INFN Sezione di Firenze, 
           Largo E. Fermi 2, I - 50125 Firenze, Italy}	
\address[mos]{Institute of Theoretical and Experimental Physics, 
   B. Cheremushkinskaya Ul, Moscow, Russian Federation}	     
\address[tai]{Institute of Physics-Academia Sinica, 
          Taipei 115, Taiwan, Republic of China}    
\address[bol]{University of Bologna - INFN Sezione di Bologna, 
   Via Irnerio,46 , I - 40126 Bologna ,Italy }  
\begin{abstract}
The Alpha Magnetic Spectrometer in a precursor version (AMS-01), was flown in June 1998 on a 51.6\mydeg~orbit and at altitudes 
ranging between 320 and 390 km, on board of the space shuttle Discovery (flight STS-91).
AMS-01 included an Aerogel Threshold \cher counter (ATC) to separate $\bar{p}$ from $e^{-}$ and $e^{+}$ from $p$, 
for momenta below 3.5 \GeVc. 
This paper presents a description of the ATC counter and reports on its
performances during the flight STS-91.
\end{abstract}
\begin{keyword}
AMS experiment, Cherenkov detectors, antiprotons
\end{keyword}
\end{frontmatter}
%
%
\section{Introduction.}
The first phase of the AMS experiment (AMS-01) was achieved on board of the space shuttle Discovery, during 10 days in June 1998. 
The main objective was to test the spectrometer's instrumentation in orbit, in preparation for the second phase that 
will take place on board of the International Space Station (ISS) for 3 to 5 years. During the shuttle flight, 
100 million events were recorded, allowing the fluxes of several particle species ($e^{\pm}$, $p^{\pm}$, He) 
to be measured~\cite{amsres}.\\
The AMS-01 detector included a permanent magnet, a Time-of-Flight scintillation counter (TOF), a silicon tracker (TRK), 
anti-coincidence scintillation counters (ACC) and an Aerogel Threshold \cher counter (ATC). A detailed description of the AMS 
spectrometer may be found in \cite{amsrep}.\\
This paper describes the ATC counter and its performances during the flight on board of Discovery.
\subsection{Role of ATC in AMS-01.}
\begin{figure}[htbp]
\begin{center}
\includegraphics[scale=0.8]{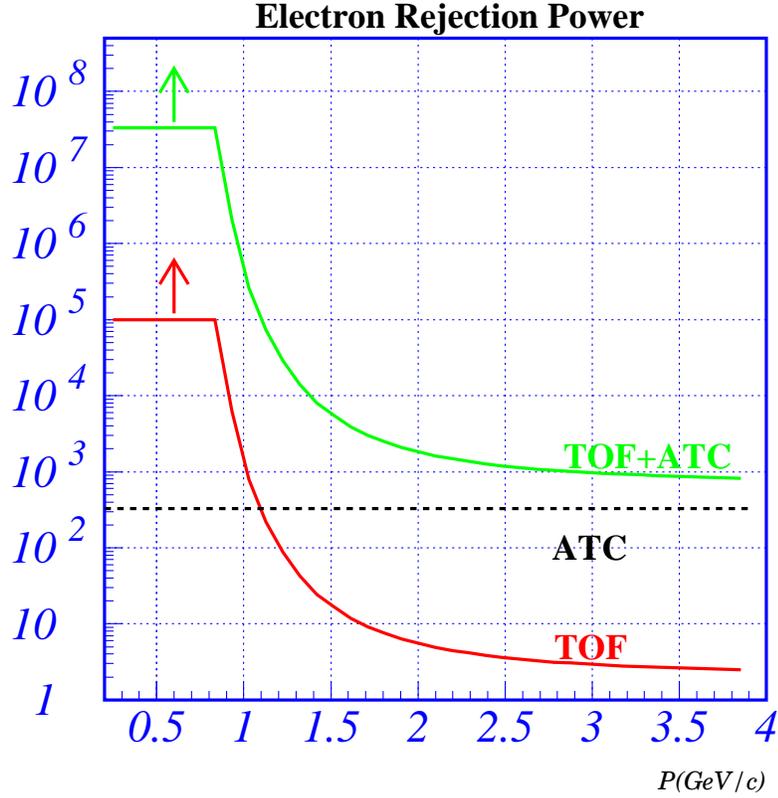}
{\noindent
\caption{AMS Rejection power as a function of P(\GeVc). The curve labelled TOF shows the rejection against 
electrons for TOF alone, for which a mass threshold has been chosen at 0.5 \GeVcc. The curve labelled 
ATC+TOF shows the rejection against electrons when taking into account both ATC and TOF. 
The improvement in the region from 1.5-2 \GeVc~to 3.5 \GeVc~is clearly seen. 
The curves have been obtained with a momentum resolution~\cite{kn:bill} ${\Delta \mathrm{P}}/\mathrm{P}=7. \, \%$ and a velocity 
resolution~\cite{kn:choumilo} 
: ${\Delta \beta}/\beta=3.3 \, \%$ together with an ATC rejection power $\mathrm{R}_{\mathrm{ATC}}=330$. 
The curves are artificially truncated for rejections above $10^5$.
\label{tofatc}}}
\end{center}
\end{figure}

One of the main purposes of the AMS Shuttle flight was to measure cosmic antiproton spectrum 
for momenta below 3.5~\GeVc~(the ATC momentum threshold). Antiproton spectrum measurement, as well as positron sprectrum, 
can be achieved by using the ATC counter :\\
$\bullet$ {\bf antiprotons :} The major background component to the $\bar{p}$ sample is expected to come from misidentified electrons.
Using the measured electron flux~\cite{amsres} and the previously measured $\bar{p}$ flux~\cite{besspbar}, 
the signal to background ratio is estimated to be : ${\bar{p}}/e^{-} \sim \!10^{-3}\!-\!10^{-2}$ 
for the considered P range.\\
While TOF counters~\cite{kn:choumilo} (${\Delta \beta }/\beta \simeq 3.3 \, \%$) allow the separation of $\bar{p}$ and $e^{-}$ 
below 1-1.5 \GeVc~(fig.~\ref{tofatc}), ATC extends this discrimination range to 3.5 \GeVc. Therefore, ${\bar{p}}/{e^{-}}$ separation 
can profit from ATC redundancy up to 1-1.5 \GeVc~and can rely
on the ATC up to 3.5 \GeVc.\\
$\bullet$ {\bf positrons : }Positrons were also an important issue for AMS-01. They had to be discriminated from a much larger proton flux, 
 with a typical ratio : ${p}/e^{+} \sim 10^{3}$. Although the ATC
design was not optimized for this selection, ${e^{+}/p}$ discrimination could be achieved by using appropriate ATC cuts~\cite{amsres}, as shown in
sec.~\ref{posi}. 
\subsection{Principle of the ATC.}
The principle of the ATC counter, used in AMS-01, is based on the \cher effect to separate $\bar{p}$ from $e^{-}$ at low energy. 
Basic relations are recalled here for the reader's convenience. 
The number of photons created by the \cher effect, in a material of refractive index $n$, is proportional to :
\begin{equation}
N_{\gamma} \propto L\times Z^{2} \times \sin^{2} \theta = L\times Z^{2} \times (1-\frac{1}{n^{2} \beta^{2}})
\label{cherequa}  
\end{equation}
\noindent where $L$ is the path length in the material, $\theta$ is 
the \cher angle, $Z$ the charge of the incoming particle and $\beta$ the particle velocity.\\
This leads to the following threshold values (in beta or momentum) : 
\begin{equation}
\beta_{thres}={1}/n~;~\mathrm{P}_{thres}=\frac{mc}{\sqrt{n^{2}-1}}
\end{equation}
\noindent
where $m$ is the mass of the particle at rest.\\
The aerogel refractive index ($n=1.035\pm 0.001$) was chosen \cite{gougas} to provide a high threshold and  
a sufficient number of photo-electrons (p.e). The corresponding thresholds, for several particle species, 
are given in the following table.
\begin{table}[htb]
\begin{center}
\begin{tabular}{|c|c|c|c|c|c|c|} \hline
Particle   & $e^{\pm}$ & $\pi^{\pm}$ & p ($\bar{p}$) & He ($\bar{He}$) \\
\hline
$\mathrm{P}_{thres}$& 1.91 \MeVc & 0.52 \GeVc & 3.51 \GeVc & 14.0 \GeVc \\
\hline
\end{tabular}
\end{center}
\end{table}
\hspace{-10mm}
\par
\noindent
Electrons and positrons in the 0.5-3.5 \GeVc~range are far above their threshold, thus giving a 
full amplitude signal in ATC. In principle, $p$ and $\bar{p}$ of momentum less than 3.5 \GeVc, 
are not expected to give any \cher signal, thus leading to ${\bar{p}}/e^{-}$ separation. 
In the following, this value of 3.5 \GeVc~will be referred to as the \cher threshold, i.e. the ATC momentum threshold for 
antiproton selection.
%
%
\section{Counter Design.}
\begin{figure}[htbp]
\begin{center}
\hspace*{-1cm}
\includegraphics[scale=0.5,angle=270]{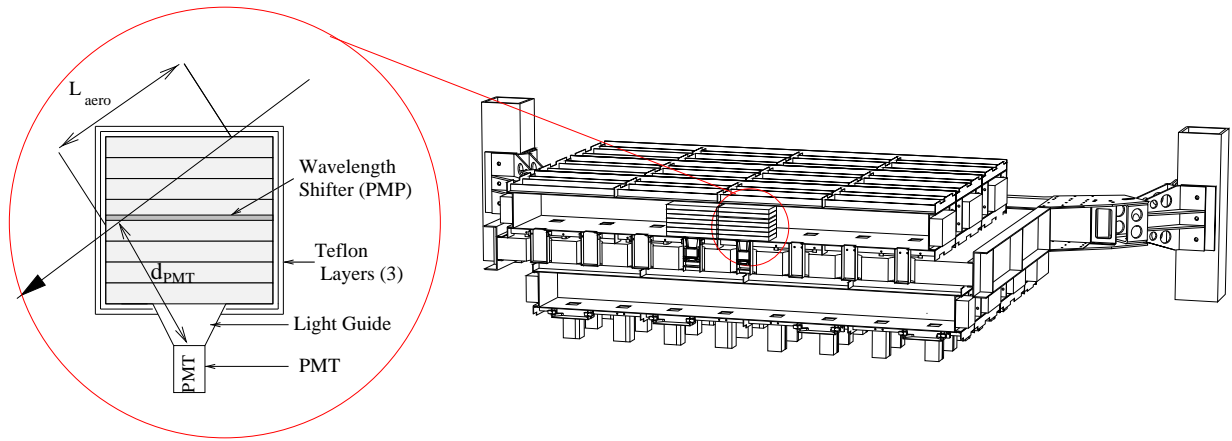}
{\noindent
\caption{View of the ATC detector design \cite{atcblois}. The 2 shifted layers (in x and y directions) can be seen together	with 
the structure that allows the ATC to be directly mounted on the unique support structure. On the ATC cell view, the 8 blocks of aerogel 
are shown together with the 3 teflon layers and the PMP wavelength shifter lying in the middle of
the cell. The impact parameter ($d_{PMT}$) used in the ATC analysis, is defined as the shortest distance between the track 
and the center of the PMT window (see sec. \ref{sec:dpm}).
\label{atcdesign}}}
\end{center}
\end{figure}

The elementary component of the ATC detector is the aerogel cell ($11\, \times \,11\, \times \,8.8 \,cm^{3}$, 
see figure \ref{atcdesign}), filled with eight $1.1\,cm$ thick aerogel blocks~\cite{kn:aero}.
The emitted photons are propagated through the aerogel material and reflected 
by three 250 \micron~teflon layers surrounding the aerogel blocks. To reach 
the photomultiplier's window (Hamamatsu R-5900), a photon crosses on average, several meters of aerogel 
and undergoes several tens diffusive reflections.\\
The limiting processes~\cite{belle} to the \cher photon detection are the Rayleigh scattering ($L_{R} \propto \lambda_{\gamma}^{4}$) and 
the absorption ($L_{abs} \propto \lambda_{\gamma}^{2}$). 
These two effects decrease with increasing photon wavelengths. 
In order to extend the counter sensitivity to the UV range, a wavelength shifter was used and placed in the middle of each 
cell (see fig.~\ref{atcdesign}). It consists of a thin layer of tedlar ($25\,\mu m$) soaked in a PMP solution 
(1-Phenyl-3-Mesityl-2-Pyrazolin) and placed into a polyethylene envelope ($50\,\mu m$), to avoid contact between PMP and the 
aerogel material. 
This allows a wavelength shift from around $300\,nm$ up to $420\,nm$. It should be noticed that the maximum efficiency 
of the R5900 photomultiplier tube is at $\lambda \sim 420\,nm$~\cite{kn:pm}.
The use of the shifter leads to an overall increase in the number of p.e estimated to be $\sim 40\%$.\\
The ATC counter was made of 168 such cells grouped in modules of $2\!\times\!8$ cells enclosed in a carbon fibre structure,  
with one special half module made of 8 cells. The modules are arranged in 2 rectangular layers 
($8\!\times\!10$ cells in the upper one and $8\!\times\!11$ cells in the lower one).
The rectangular shape of the layers is designed to maximize the acceptance, and the second layers is shifted by 
half of a module width (fig.~\ref{atcdesign}) to minimize the loss of tracks passing between cells. 
The 2 layers are bolted respectively above and below a $5\,cm$ thick honeycomb plate glued inside a 
frame of aluminium mounted on the unique support structure (USS) by four brackets.

\begin{figure}[htbp]
\begin{center}
\includegraphics[scale=1.4]{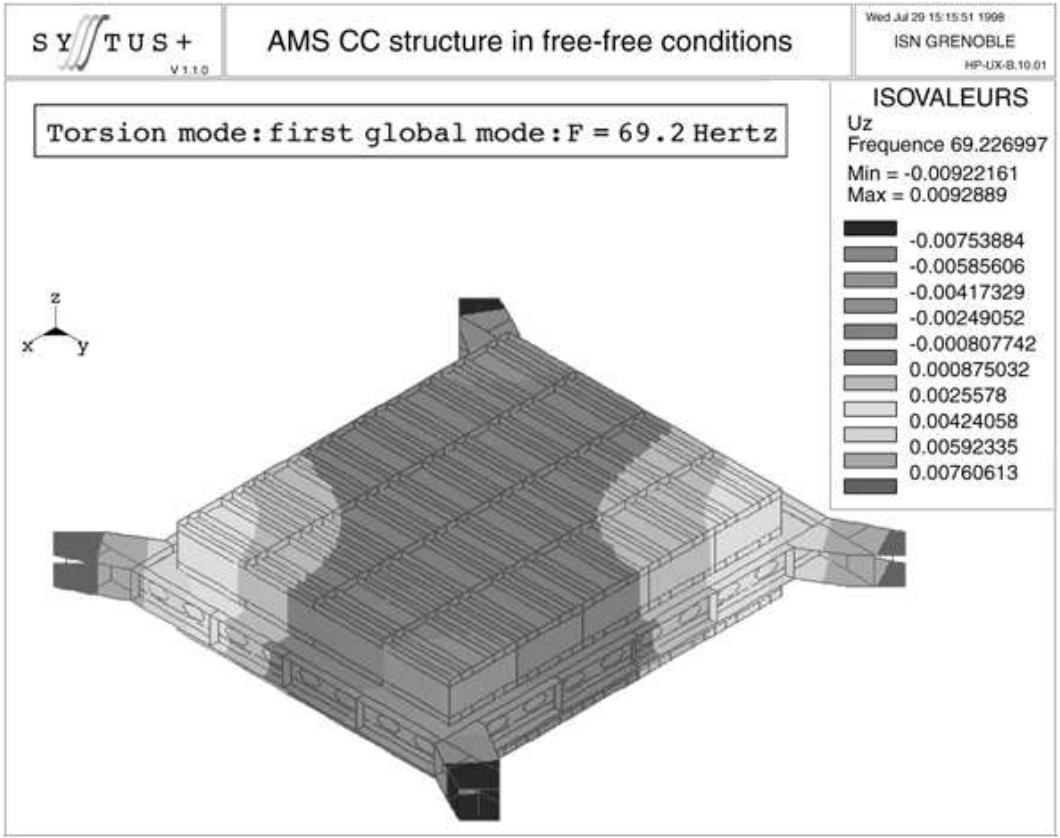}
{\noindent
\caption{Finite element calculation for the estimation of the dynamical response of the ATC. The lowest 
eigenfrequency is found at 69.2 Hz.
\label{systus}}}
\end{center}
\end{figure}

The mechanical design of the ATC was an important issue of the AMS-01
experiment, due to the fact that the ATC was mounted directly to the USS 
independently of the rest of the detector.\\
The minimization of the total mass of the ATC counter was a crucial requirement
to be fulfilled. The final mass was 120 $kg$.\\
The safety margins were carefully controlled by NASA.
A finite element model (FEM) using the SYSTUS software~\cite{ref:systus} was developed to estimate the 
dynamical response of the mechanical structure to the extreme conditions of 
the shuttle launch. The lowest eigenfrequency of the ATC structure had to be
as far as possible from the eigenmode of the USS being around 
10 Hz. The FEM allowed to validate the whole mechanical design, showing the 
lowest eigenfrequency in the free-free configuration at 69.2 Hz (fig.~\ref{systus}). In the 
constrained configuration with the four corners coupled to the USS, the FEM 
gave a value of 40 Hz well above the 10 Hz range of the USS. The value
of the eigenfrequency in the free-free configuration was confirmed by a 
"smart hammer" test performed \cite{entre} just before the 
integration of the ATC counter with the rest of the detector. The measured value
was  66.9 Hz for the torsion mode and 75.8 Hz for the drum mode.

%
%
\section{Electronics of the AMS-01 Aerogel Threshold \cher counter.}
\subsection{Electronics principle.}
\label{elecprinc}
\begin{figure}[htbp]
\begin{center}
\includegraphics[scale=0.8,angle=270]{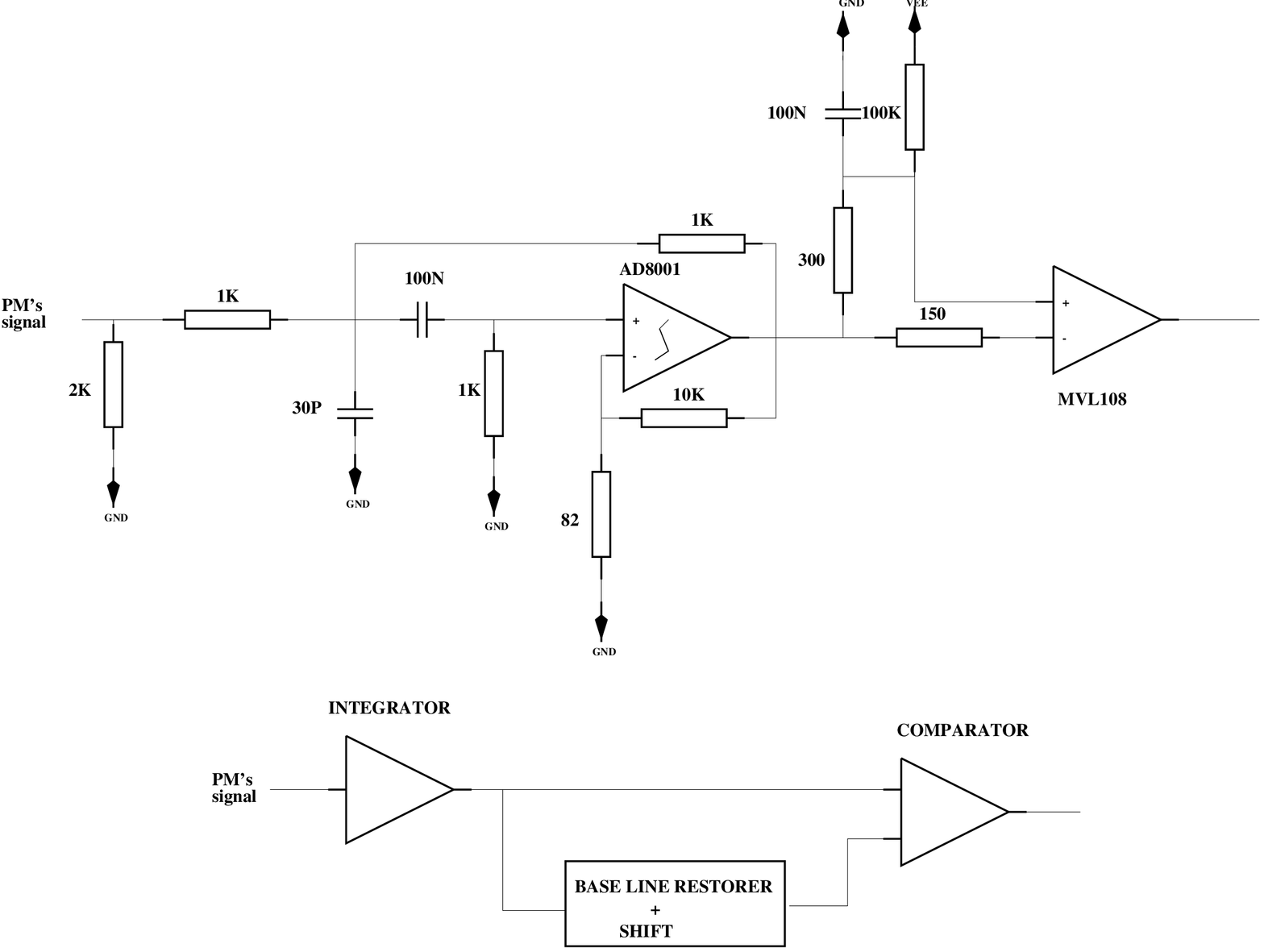}
{\noindent
\caption{Scheme of the ATC counter electronics.}
\label{fig:elec}}
\end{center}
\end{figure}

The electronics used for the ATC was derived from the one designed for the Time of Flight counter~\cite{tofal} of AMS-01.\\
An analog board integrates the PMT signal and compares it to a threshold. This threshold is fixed at a 
level above the photomultiplier and electronic noises. This way, the input signal 
(about ten nanoseconds wide) is transformed into a logical signal (in the hundreds nanoseconds of range) whose length is proportional to the 
logarithm of the collected charge. This logical signal is then transferred to 
a digital card, and converted with a TDC (Time to Digital converter), as for the TOF signal.\\ 
The design principles of the TOF electronics were kept whilst adapting the impedances and dynamic range and 
including a base line restorer 
at the integrator output. Thanks to this design, the integrator offsets are automatically compensated and do not 
need to be calibrated. On the other hand, the use of TDC means the original charge signal can be corrupted by 
after-pulses~\cite{after} happening during the integration time ($\sim$ 200 nsec).\\
The scheme of the electronics is presented in fig.~\ref{fig:elec}.
\subsection{Electronics channels.}
As mentioned above, each cell of the ATC was read by one photomultiplier tube (PMT). The PMTs of 2 consecutive diagonal cells 
in a module are grouped per electronic channel. There was 84 such channels. 
An analog board (SBBC), located close to the PMT, handled 2 channels. A logical board (SFEC) handled 14 channels plus 2 
trigger channels. 
For redundancy, there was two high voltage boards (SHVC boards) for each module\\
\subsection{Tests and Space qualification.}
The design of the AMS electronics has followed some of the space qualification rules. 
The choice of the electronic components has followed the rules for a manned flight.\\
The printed circuit boards have been designed and manufactured according to precise space constraints : strip width, pad size, board coating 
... as described in the ESA standards~\cite{esa}. Connectors and cables match ``standard space'' specifications. 
The PMT high voltage supply have a dual redundant architecture.\\
In addition to common electrical tests, the analog and high voltage units passed thermal and vacuum tests. The digital boards passed thermal 
and vibration tests under vacuum.     
\subsection{ATC Calibration}
The calibration of the ATC counter has provided, for each cell, the coefficients 
of the expression relating the number of photo-electrons to the time measurement provided
by the SFEC cards. All PMTs inside the same module were supplied with the same high voltage. The pre-selection allowed to limit the 
gain dispersion to less than 30 \%.
Formula (3) was  used to describe the electronic response. It is a good approximation on the small dynamic range of the ATC signal. 
It involves the PMT gains and the electronics characteristics:    
\begin{equation}    
                 Q_{e}(i)=A \times g(i)\times e^{\frac{t+C_{2}(j)}{C_{3}}}
\end{equation}
\noindent where :
%
%
%
%
%
%
%
%
%
$Q_{e}(i)$ are the detected charges in photo-electron unit, A is the overall normalization factor (same value for all 168 cells), 
g(i) are the relative gains (average value equal to 1) for each PMT $i$, $C_{3}$ is the parameter governing the logarithmic behaviour of the 
front-end electronics and $C_{2}(j)$  are the time "pedestals", relative to each electronic channel $j$.

\noindent	
The coefficient $C_{3}$ of the exponential could not be correctly measured card by
card before the ATC mounting because its value depends on the final
ATC cabling and PMT input capacitances on the amplifiers. Measurements
on the ATC after the flight have shown that, for all amplifiers,  $C_{3}$ was equal to $220 \;ns$ within $\pm 2\%$.\\
The $C_{2}$ coefficients are taken from the raw data time distributions. 
The final coefficients were obtained, from the flight data themselves. Selecting protons below \cher threshold 
allow to measure the peak at one p.e (coming from the residual scintillation light of 0.5 p.e on average per cell) 
with high statistics. 
The main problem was that 25-30$\%$ of the R-5900 PMTs were not giving a nice single p.e distribution. Nevertheless, 
the high number of detected protons has permitted the 168 gains to be equalized.\\
As described below (see sec.~\ref{npebeta}), the ATC luminosity (number of p.e for $\beta \simeq 1.$) is obtained using high $\beta$ protons from the flight STS-91.
\subsubsection{Front-end electronics thresholds}
\begin{figure}[htbp]
\begin{center}
\includegraphics[scale=0.8]{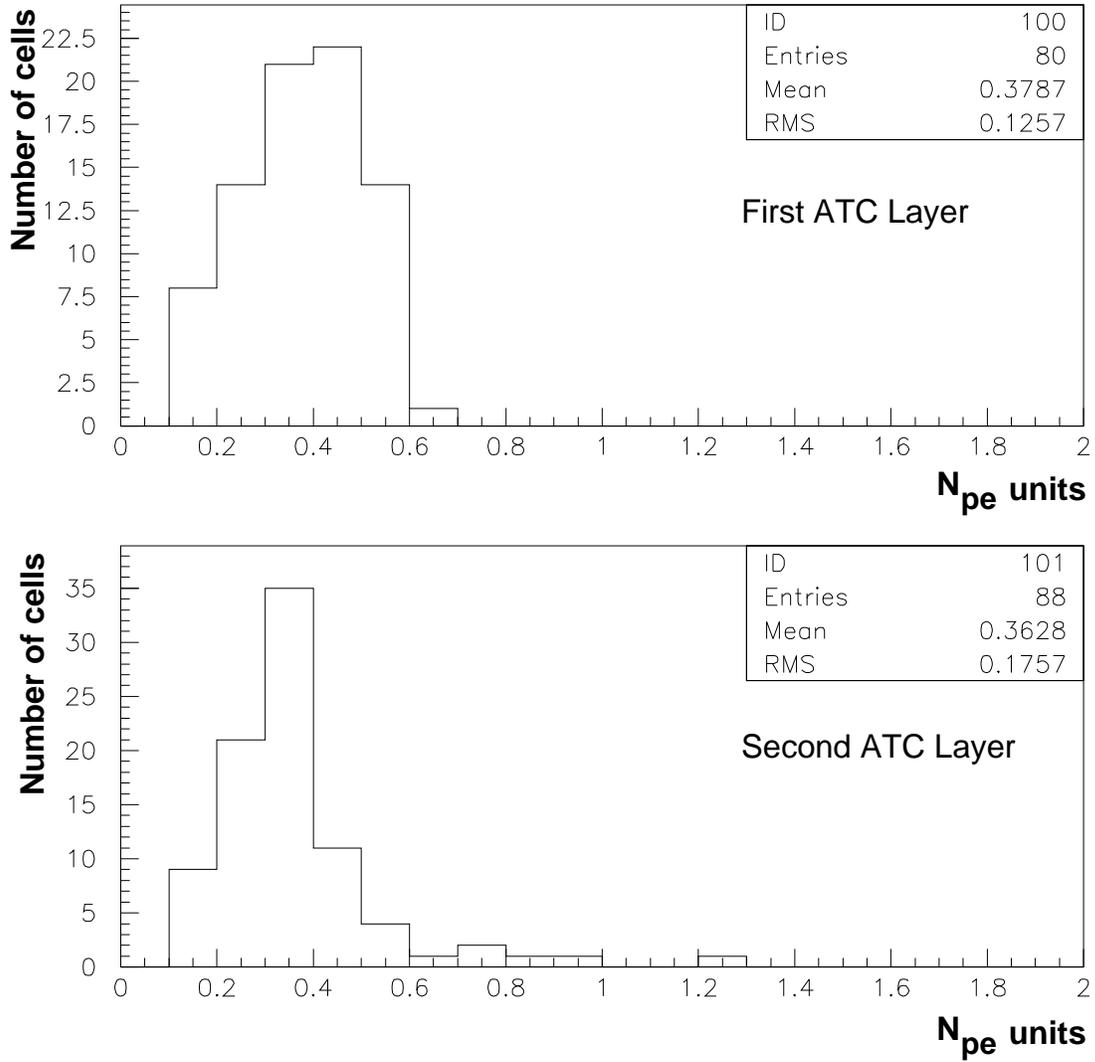}
{\noindent
\caption{ Distribution of the 168 thresholds in photo-electron units, for the upper and lower ATC plane. 
The spreading reflects the dispersion of the PMT gains. Note that the same value of high voltage was 
used for a group of 16 pre-selected PMTs.
\label{jf2}}}
\end{center}
\end{figure}

Each front-end electronic channel has a threshold. Although they were tested to
be very similar from channel to channel (dispersion below 10$\%$), the differences
of PMT gains induced a dispersion over the thresholds calculated in photo-electron
units. Fig.~\ref{jf2} shows the distribution of the 168 thresholds. The mean
value is seen to be 0.37 p.e. This threshold was taken into account, using a Poisson distribution, 
for estimating the number of p.e for $\beta \simeq 1$. The correction is 15$\%$ on average.
%
%
\section{Detector Performances.} 
In this section, the ATC performances during the flight STS-91 is discussed, namely the average 
number of photo-electrons for $\beta \simeq 1$ particle (sec. \ref{npebeta}), the observed ageing problem (sec.~\ref{sec:age}), the signal for protons and heliums 
(sec. \ref{zdep}), the dependence of the value of $n_{p.e}$ on the impact parameter (sec. \ref{sec:dpm}), 
as well as a cross-check of the refractive index 
of the aerogel medium (sec. \ref{sec:index}).
\subsection{Absolute number of photo-electrons ($n_{p.e}$) for $\beta \simeq 1$ particles.}
\label{npebeta}
The method used \cite{atcint} to evaluate $n_{p.e}$ in each cell crossed by a $\beta \simeq 1$
particle (electron or proton above the \cher threshold) consists of comparing the number of "no signals" 
in the cell with the number expected by the Poisson distribution for a given average $n_{p.e}$. It must be
outlined that this method is the closest to the ATC operation.
By using particles selected as high energy ones during the flight STS-91 (protons with $P \geq 15$ \GeVc, see section \ref{pbar} 
for a definition of this control sample), by ensuring that the particle is really 
crossing the cell (track
qualities selection) and by taking into account the charge thresholds of the electronics, the
following average $n_{p.e}$ is obtained :
\begin{displaymath}
\left\{ \begin{array}{ll}
n_{p.e}=2.9 \mathrm{,\: for\: the\: upper\: plane}\\
n_{p.e}=3.3  \mathrm{,\: for\: the\: lower\: plane}
\end{array} \right.
\end{displaymath}
\noindent			
After correcting for electronic thresholds and the average $\beta$ effect 
of the proton sample, these values become :
\begin{displaymath}
\left\{ \begin{array}{ll}
n_{p.e}=3.51\pm0.02 \mathrm{,\: for\: the\: upper\: plane}\\
n_{p.e}=4.02\pm0.02 \mathrm{,\: for\: the\: lower\: plane}
\end{array} \right.
\end{displaymath}
\subsection{ATC Monitoring and the ageing problem.}
\label{sec:age}
The ATC has suffered a fast degradation with time of its \cher yield. The average $n_{p.e}$ by cell (for $\beta \simeq 1$) was about 
5 p.e/cell in November 97. During the flight, this value had decreased to 3.1 p.e/cell
and at the November 98 CERN test beam, it was down to 1.5 p.e/cell, which corresponds to an equivalent 
life time of about 300 days. During the same period, a
reference cell at room temperature has decreased with a much larger life time of 1044 days. 
A study of the effects of various materials on the aerogel was performed before and after the STS-91 flight~\cite{age1,age2}. It was found 
that the aerogel is insensitive to the presence of water vapours. On the contrary, PMP deposited directly on the surface of the aerogel induces 
a fast degradation of the light transmission, specially for wavelengths in the blue range ($465\;nm$). This evolution was stopped by cooling the aerogel,
which indicates clearly that the effect is due to a chemical reaction. The PMP in the cell was thus isolated from the aerogel by a plastic bag which 
was efficient enough, as demonstrated by the reference cell. However, the aerogel was not isolated from possible organic vapours from the black RTV used
for the light tightening of the ATC. This is the most probable source of the ageing problem observed\footnote{ Several studies~\cite{age1,age2} 
have been made to investigate this ageing issue in the visible range, using the same aerogel material.}.\\ 
The ATC counter was continuously monitored during the flight STS-91. Only a few cells showed some electronic problems and
were discarded from the analysis \cite{atcint}. The effective acceptance was eventually 93 \% of the geometrical one.
\subsection{ATC signal for protons and heliums.}
\label{zdep}
\begin{figure}[htbp]
\begin{center}
\includegraphics[scale=0.8]{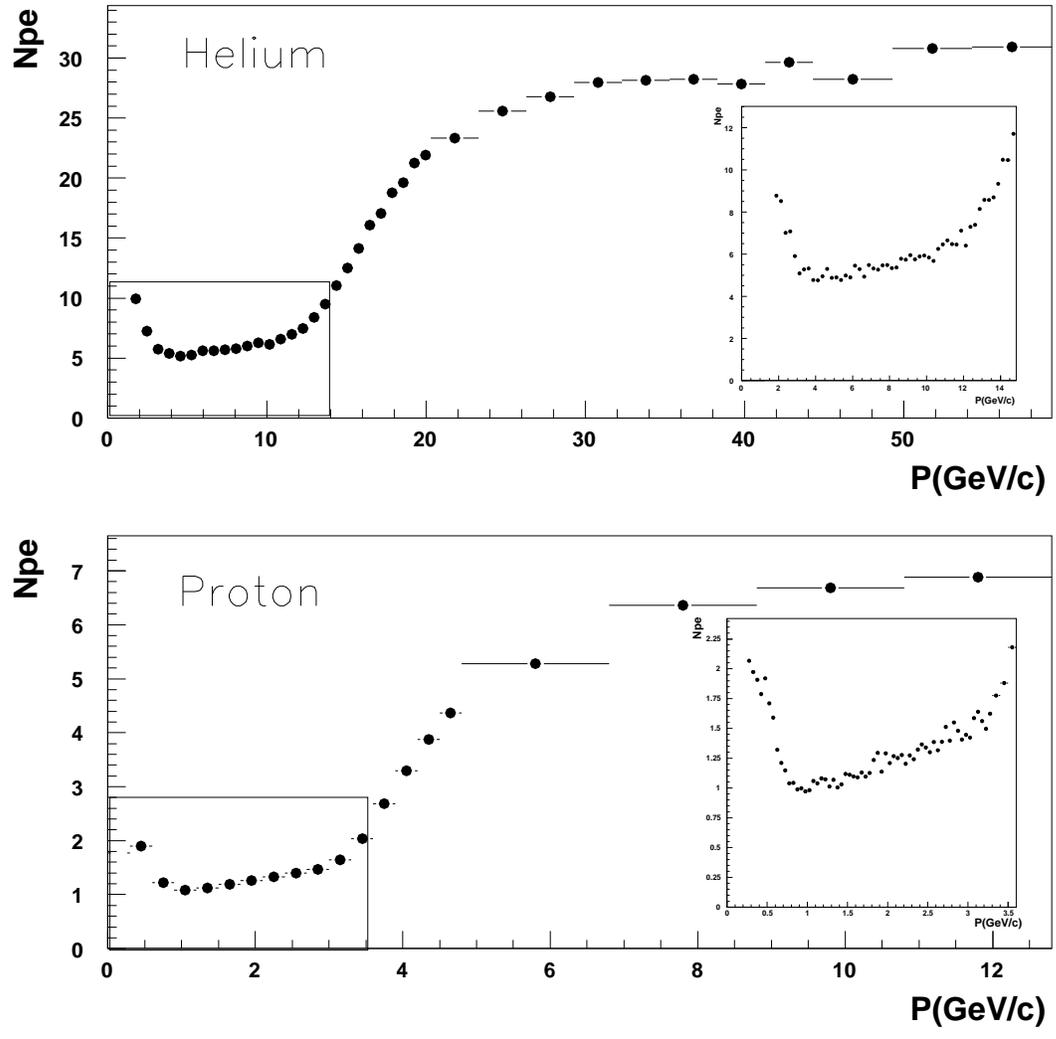}
{\noindent
\caption{ Helium and proton spectra. The upper (lower) figure shows $n^{tot}_{p.e}$ vs
P(\GeVc) for helium (proton). A residual light (scintillation and \cher effect out of the aerogel) can be observed 
below the \cher threshold, see text.}
\label{z2}}
\end{center}
\end{figure}

Figure~\ref{z2} shows the number of photo-electrons ($n^{tot}_{p.e}$) as a function of P(\GeVc), for particles identified by AMS
 as helium (upper part) and proton (lower part). As expected, the number of photo-electrons is proportional 
 (eq.~\ref{cherequa}) to the square of the particle's charge.
Above the \cher threshold, the number of p.e follows the expected dependence.
\begin{equation}
n_{p.e} \propto (1+{m^{2}}/P^{2})
\end{equation} 
Far above threshold, where the \cher yield saturates, the signal ratio for helium and proton particles is in good
agreement with the expected factor of 4.\\
A residual light can be observed (see fig.~\ref{z2}) below the \cher threshold. It can be
evaluated to be $\sim 1 \,p.e$ for protons, summed over the 2 ATC layers. This is due to $\delta$-rays, to
~\cher effect in the wavelength shifter component, and to scintillation. 
Below $\sim 1.$ \GeVc~an increase in the residual light due to scintillation 
is observed (see detailed view in fig.~\ref{z2}). The residual light is also slightly increasing between 1 \GeVc~and 3 \GeVc. 
This may be explained\footnote{ We estimate the threshold for the ~\cher effect in these materials 
to be $\sim \!1.$ \GeVc~. At 2.7 \GeVc~the $n_{p.e}$ will be 10 \% of the maximum value.} by $\delta$-rays, and the ~\cher effect 
in the polyethylene and teflon layers. This effect will cause the proton rejection power (for $e^{+}$ selection), see equation 
\ref{eq:rej}, to decrease between 1 \GeVc~and 3 \GeVc~(see fig.~\ref{eplussel}). On the other hand it will only slightly affect the
$\bar{p}$ efficiency (sec. \ref{pbar}).
\subsection{Distance of track to the PMT.}
\label{sec:dpm}
The shortest distance between the track and the center of the PMT window (see figure \ref{atcdesign}), 
is defined as the impact parameter ($d_{PMT}$). It turns out that this variable is strongly correlated with the cell signal.\\
In fact, the number of collected photo-electrons is expected to increase with decreasing impact parameter 
for several reasons :
\begin{itemize}
  {
\item The probability of photons entering directly into the PMT window (without any reflections) is increased, which 
limits losses due to reflections.
\item  The path length of photons in aerogel is shorter on average leading to less absorption.
\item For a very low impact parameter ($d_{PMT} \leq 1\,cm$), the particle is expected to produce a large number of 
photons from the \cher effect in the PMT window.}
\end{itemize}
Figure \ref{dpm} shows the number of p.e as a function of the impact parameter distance squared. 
The signal enhancement at low impact parameters is clearly visible and is observed to be $\sim 6 \,p.e$ per layer. 
As it will be shown in section \ref{posi}, this variable is used to enhance the 
proton rejection when selecting positrons.
\subsection{Refractive Index Evaluation.}
\label{sec:index}
\begin{figure}[htbp]
\begin{center}
\includegraphics[scale=0.8]{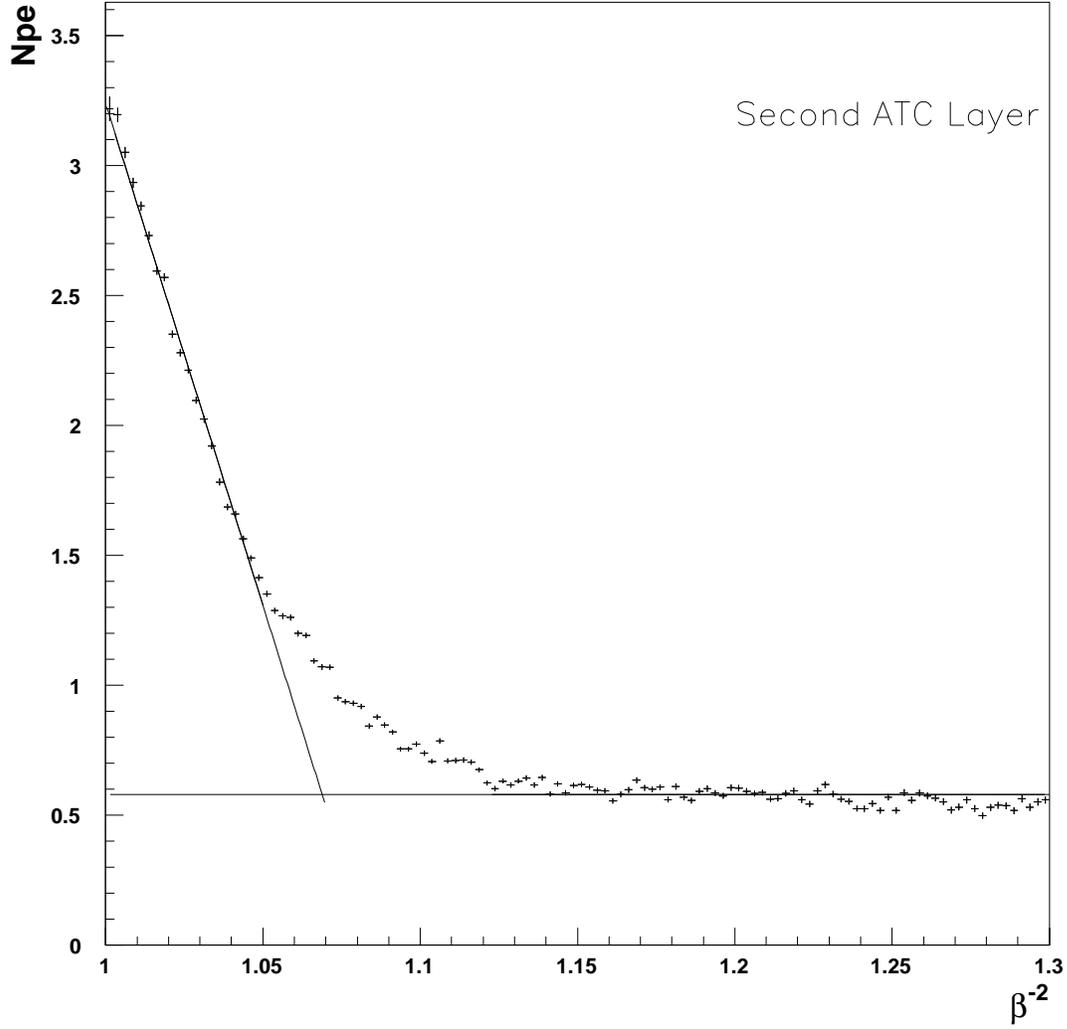}
{\noindent
\caption{ Refractive index evaluation. Number of photo-electrons ($n_{p.e}$) in the second ATC layer against ${1}/\beta^{2}$. 
For ${1}/\beta^{2}$ greater than 1.1 (corresponding to low momentum), the residual light can be observed 
when $n_{p.e}$ reaches a minimum.
Below 1.1, $n_{p.e}$ is increasing with decreasing ${1}/\beta^{2}$. The extrapolation of the observed threshold leads to an evaluation of the refractive
index.}
\label{nbeta2}}
\end{center}
\end{figure}

Using the flight STS-91 data, it is possible to evaluate the refractive index of the aerogel. 
Above the \cher threshold, $n_{p.e}$ is expected to be a linear function of ${1}/\beta^{2}$, as shown in equation 1. 
Figure \ref{nbeta2} shows the number of p.e as a function of ${1}/\beta^{2}$, for the second ATC layer.
Using the extrapolation of the background (residual light), 
one can extrapolate the observed threshold and thus evaluate the value of $n$.
\begin{equation}
n^{2}={1}/\beta_{thres}^{2} \simeq 1.07 \, \rightarrow \, n \simeq 1.0344
\end{equation}
Neglecting index dispersion, this value is in good agreement with the refractive index measured \cite{gougas} before the flight 
($n=1.035 \pm 0.001$) and the value given by the manufacturer~\cite{kn:aero} ($n$=1.035).
%
%
%
%
\section{Particle selection with ATC.}
\subsection{${\bar{p}}/e^{-}$ discrimination}
\label{pbar}
\begin{figure}[htbp]
\begin{center}
\includegraphics[scale=0.8]{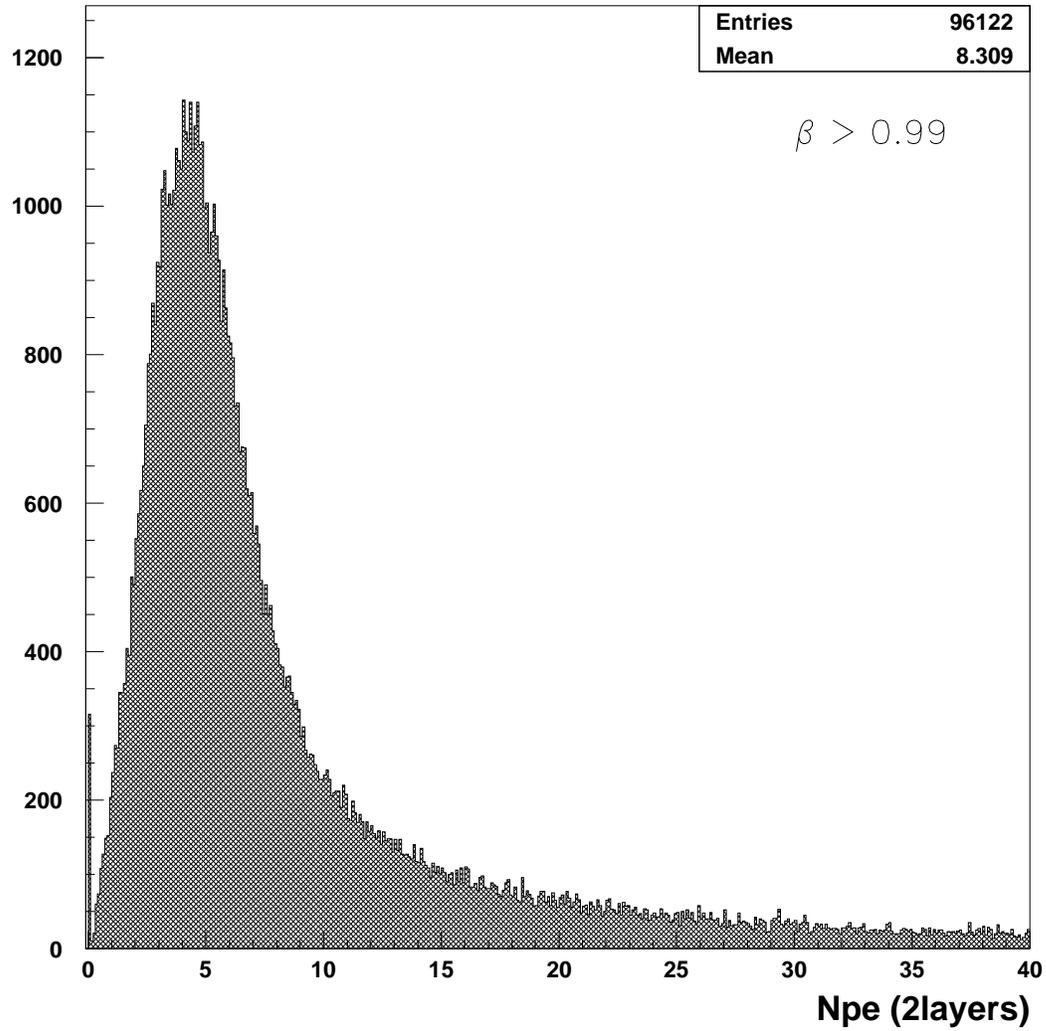}
{\noindent
\caption{ Distribution of $n_{p.e}$ from particles above the \cher threshold (with $P \geq 15 $ \GeVc~and $\beta \geq 0.99$ cuts)
 selected near magnetic equator. This is similar to the behaviour of $e^{-}$, which is always above the \cher threshold. 
 This sample is used to evaluate ATC rejection against electrons.}
\label{npeelec}}
\end{center}
\end{figure}

\begin{figure}[htbp]
\begin{center}
\includegraphics[scale=0.8]{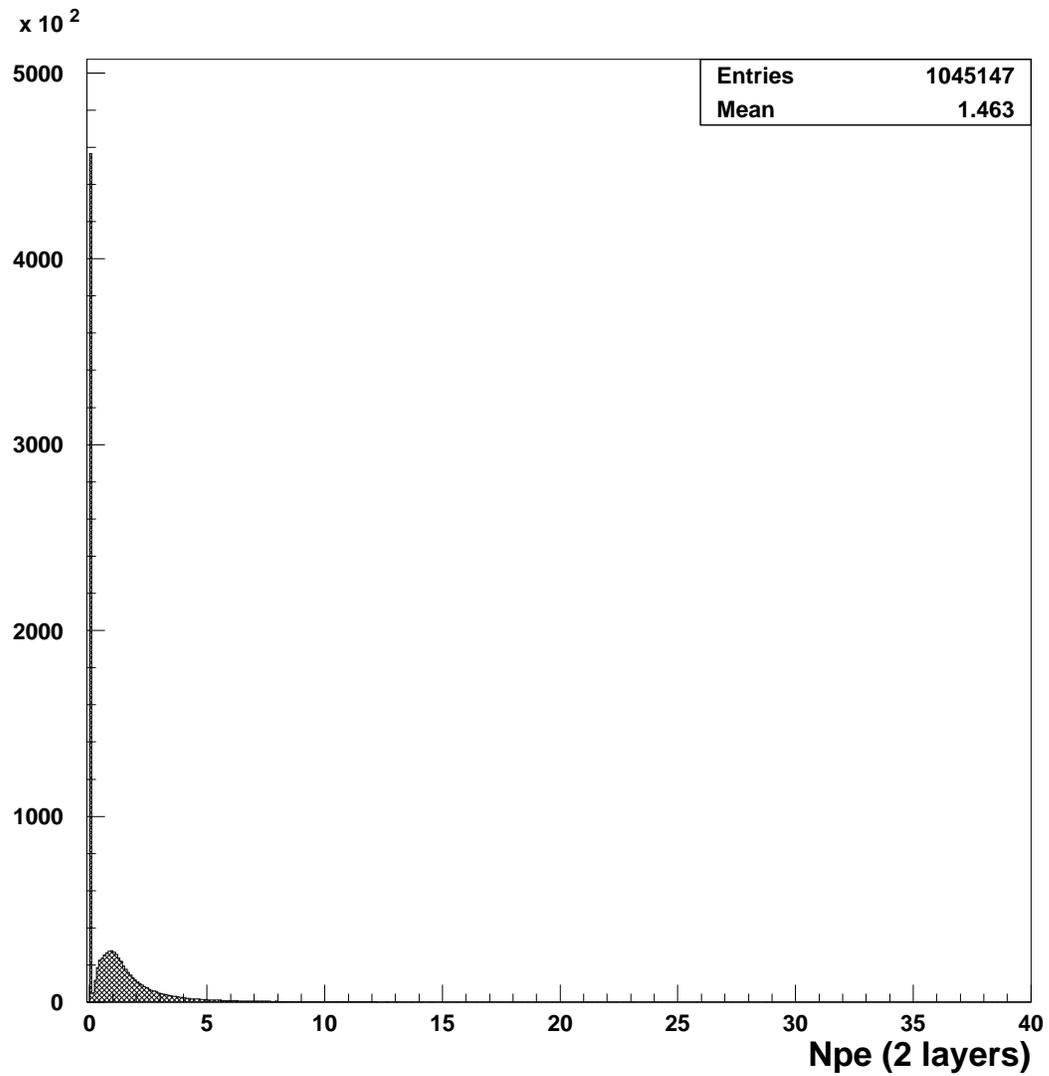}
{\noindent
\caption{ Distribution of $n_{p.e}$ from particles below the \cher threshold (low energy protons selected with $P \leq 3.5$ \GeVc, 
$\beta \leq 0.97$ and $0.6 \leq M \leq1.2$ \GeVcc). Low energy $\bar{p}$ have the same behaviour in the ATC.}
\label{npepro}}
\end{center}
\end{figure}

\begin{figure}[htbp]
\begin{center}
\includegraphics[scale=0.8]{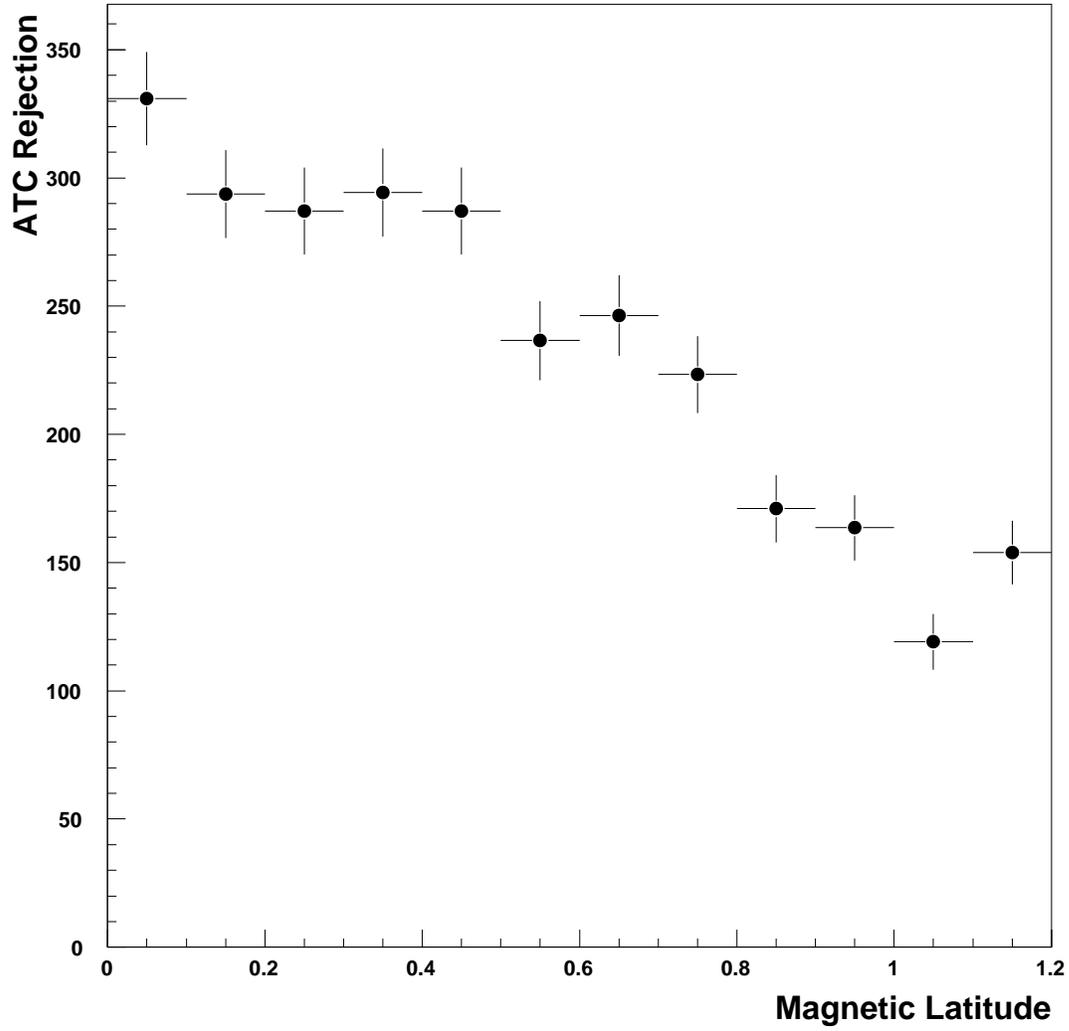}
{\noindent
\caption{ ATC Rejection ($n_{p.e}\leq 0.15)$ against electrons as a function of magnetic latitude. 
The selected sample of $\beta \simeq 1$ particles shows a higher purity for low magnetic latitude (near equator)
 as there the geomagnetic cutoff is maximum.}
\label{rejthetamag}}
\end{center}
\end{figure}

\begin{figure}[htbp]
\begin{center}
\includegraphics[scale=0.8]{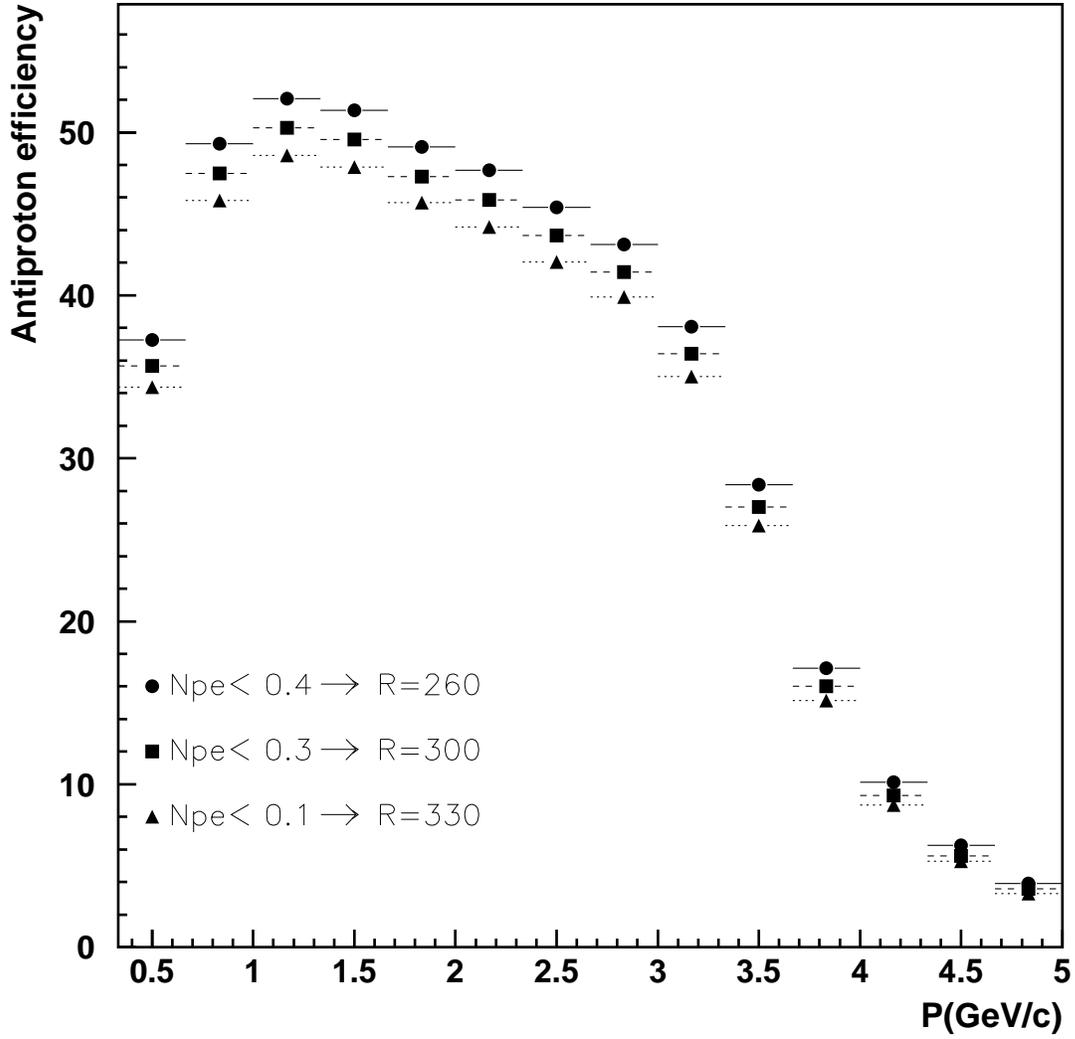}
{\noindent
\caption{ ATC selection efficiency for protons (antiprotons) as a function of P(\GeVc) for different cuts on $n_{p.e}$. 
The electron rejection (R) power is also indicated for each cut.}
\label{atceff}}
\end{center}
\end{figure}

The discrimination of $\bar{p}$ from $e^{-}$ background is obtained using two offline conditions. Firstly the particle must have 
crossed the 2 ATC layers. This requirement leads to an overall geometrical efficiency of $72\%$. Secondly the particle must have 
not produced any signal in the ATC, leading to antiproton detection efficiency ($\epsilon_{\bar{p}}$) and electron 
rejection ($R_{e^{-}}$) as discussed below.\\
Two control samples are being used to estimate $\epsilon_{\bar{p}}$ and $R_{e^{-}}$. 
Particles above the \cher threshold (high energy protons near equator, with P $\geq$ 15 \GeVc~and $\beta \geq 0.99$) 
will have the same \cher yield as electrons, whereas a sample of particles below the \cher threshold (protons with low $\beta$) 
is used to evaluate low energy antiproton \cher yield.
Figures \ref{npeelec} and \ref{npepro} show the distributions of $n_{p.e}$ for these two samples of particles.\\
In figure \ref{rejthetamag}, the ATC rejection power is presented as a function of the magnetic latitude ($\theta_{mag}$).
It can be seen that the rejection is better near the equator, indicating that the sample of high energy particles is 
less contaminated since the geomagnetic cutoff~\cite{cutoff} discards low energy cosmic particles. 
Thus we take advantage of the geomagnetic cutoff to select a sample of high energy particles with a small low energy component, 
by imposing $\theta_{mag} \leq 0.5 \,rad$, 
where $\theta_{mag}$ is the magnetic latitude evaluated with a shifted dipole model.\\
As shown in figure \ref{npeelec}, most particles above the \cher threshold give $\sim 6 \,p.e$ (summed over the 2 ATC layers). 
A tail of higher numbers of p.e, due to after-pulses in the PMT, can be observed.
On the other hand, some percentage of $e^{-}$ lead to a low signal in ATC. This is due to statistical fluctuations and is related to 
the $e^{-}$ rejection. For a given cut on $n_{p.e}$, the $e^{-}$ rejection power ($R_{e^{-}}$) is defined as :
\begin{equation}
  R_{e^{-}}(n_{cut})=~{\mathcal{N}}_{e^{-}}/{\mathcal{N}}_{e^{-}}(n_{p.e} \leq n_{cut})
\label{eq:rej}  
\end{equation}  
\\
Particles below the \cher threshold are selected as protons of momentum less than 3.5 \GeVc~and $\beta$ less than 0.97.
It can be noticed on fig.~\ref{npepro} that most of the low energy protons do not give any signal. The residual light effect can be observed around 1 p.e with a tail produced by 
$\delta -$rays, scintillation  and after-pulses.
For a given cut on p.e number, the $\bar{p}$ detection efficiency is defined as :
\begin{equation}
\epsilon_{\bar{p}}(n_{cut})={\mathcal{N}}_{\bar{p}}(n_{p.e} \leq n_{cut})/{\mathcal{N}}_{\bar{p}}
\end{equation}  
Figure \ref{atceff} shows the antiproton efficiency as a function of P(\GeVc) for different cuts on $n_{p.e}$. 
The electron rejection power is also shown for each cut.
Based on figure \ref{jf2} (distribution of electronic thresholds), an ATC cut for 
antiproton selection can be chosen as :
\begin{equation}
n_{p.e} \le 0.15~p.e
\end{equation}
For this cut, the rejection is $R_{e^{-}}\simeq 330$ and the efficiency is up to $\epsilon_{\bar{p}} \simeq
48\%$, depending 
on the momentum (see fig.~\ref{atceff}).\\
As a conclusion, it can be said that the ATC provides a rejection power of $\sim 330$ against electrons.
This information is combined with Tracker and TOF measurements to determine $\bar{p}/p$ up to 2-2.5 \GeVc. Above
this momentum, and up to 3.5 \GeVc, the evaluation of $\bar{p}/p$ is based on the single ATC data.
\subsection{$e^{+}$ selection with ATC.}
\label{posi}
\begin{figure}[htbp]
\begin{center}
\includegraphics[scale=0.8]{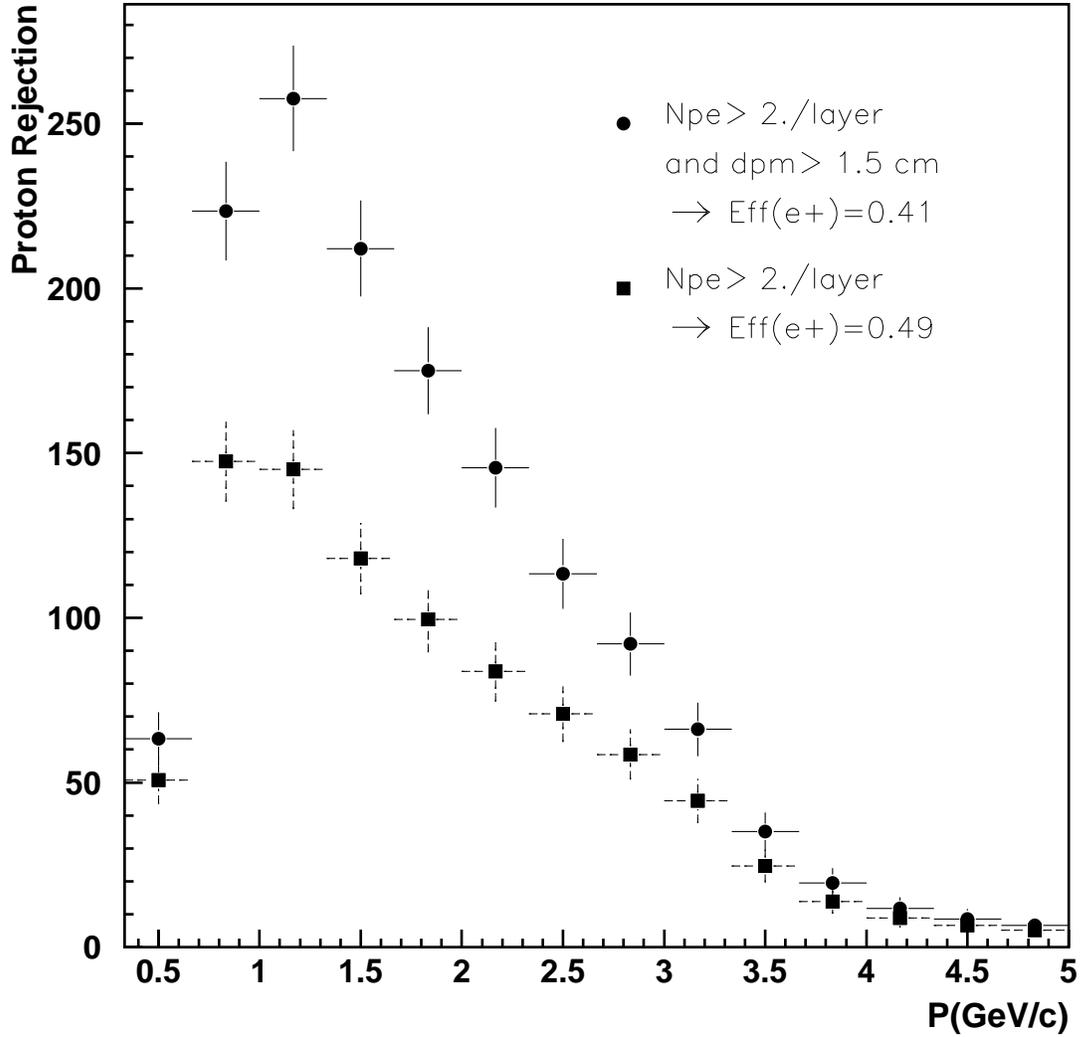}
{\noindent
\caption{ ATC positron selection : Proton rejection as a function of P(\GeVc) for the 2 sets of selection criteria.
The positron efficiency ($\epsilon$) is also indicated for each cut.
It can be noticed that in both cases the ATC proton rejection reaches a maximum near 1 \GeVc , where the ATC proton signal is at minimum
(see fig.~\ref{z2})}
\label{eplussel}}
\end{center}
\end{figure}

The ATC counter has also been used to discriminate $e^{+}$ from protons, although its design was not optimized for such a discrimination. 
Such a selection allows to extend $e^{+}/(e^{-}+e^{+})$ measurement up to 3.5 \GeVc.\\
Positrons are expected to deposit a \cher signal in each ATC layer. On the other hand, protons below  
threshold do not give any \cher signal in the aerogel material, but will potentially contaminate the $e^{+}$
selection due to both non-\cher signal of every cell (scintillation, $\delta$-rays) and to the \cher radiation in other materials and 
in the PMT-windows. 
Furthermore these physical signals can be artificially stressed by after-pulse effects on the  PMT (see sec. \ref{elecprinc}).\\
Positrons are then selected by requiring a path length in aerogel greater than $8\,cm$/layer (fig.~\ref{atcdesign}) and a number 
of photo-electrons greater than $2\,p.e$/layer.\\
The efficiency so obtained is $\epsilon_{e^{+}} \sim$ 45\% with a proton rejection up to $R_{p} \sim 150$.\\
The proton contamination coming from particles passing close to the PMT can be reduced with an appropriate cut on the impact parameter. 
Requiring the minimal impact parameter to be greater than $1.5\,cm$, sets the selection efficiency 
to $\epsilon_{e^{+}} \sim$ 41 \% and the proton rejection up to $R_{p} \sim 260$ (see fig.~\ref{eplussel}).\\
The proton rejection as a function of P(\GeVc) is shown in figure \ref{eplussel}. It has a maximum value at 1 \GeVc, where the ATC
proton signal is minimum (see fig.~\ref{z2}).\\
The ATC counter provides an efficient discrimination between $e^{+}$ and $p$ background, in the 0.5-3.5 
\mbox{$ {\mathrm{GeV}}/c\,$} range, which has been used for the AMS-01 lepton analysis \cite{amsres}.
%
%
%
%
\section{Conclusion.}
The general behaviour of the AMS apparatus was satisfactory during its first test flight on board the space
shuttle Discovery \cite{amsrep}. The ATC counter allowed 
${\bar{p}}/e^{-}$ separation with a rejection of 330 and an efficiency up to $\sim 48 \%$, extending 
the ${\bar{p}}/e^{-}$ separation range up to 3.5 \GeVc. As a secondary
result, it has been used, with appropriate cuts, to separate $e^{+}$ from protons with a
rejection up to 260 and an efficiency of $\epsilon_{e^{+}} \simeq$ 41 \%.
\newpage    
\listoffigures
%
%

%
%
%
\newpage
\newpage
\newpage
\newpage
\newpage
\newpage
\newpage
\begin{figure}[htbp]
\begin{center}
\includegraphics[scale=0.8]{figures/dpm.epsi}
{\noindent
\caption{ Distance to PMT. The upper (lower) curve shows $n_{p.e}$ as a function of the square of the impact 
parameter ($d_{PMT}^{2}$, in cm${^2}$) for the first (second) ATC layer.}
\label{dpm}}
\end{center}
\end{figure}
\newpage
\newpage
\newpage
\newpage
\newpage     
\newpage     
\end{document}